\documentclass[sigconf, nonacm]{acmart}

\AtBeginDocument{%
  }
\newcommand{\hide}[1]{}

\usepackage{amsmath,amsfonts}
\usepackage{algorithm}
\usepackage{algorithmicx}
\usepackage{algpseudocode}
\usepackage{graphicx}
\usepackage{textcomp}
\usepackage{xcolor}
\usepackage{multirow}
\usepackage{subfigure}
\usepackage{xurl}
\usepackage{xspace}
\newcommand{\name}{QASM\xspace}

\setcopyright{none}
\renewcommand\footnotetextcopyrightpermission[1]{}
\begin{document}
\title{QASM: A Novel Framework for QUIC-Aware Stateful Middleboxes}

\author{Hari Hara Sudhan Selvam and Sameer G. Kulkarni}
\affiliation{
  \institution{Indian Institute of Technology Gandhinagar}
  \state{Gujarat}
  \country{India}
}
\renewcommand{\shortauthors}{Hari Hara Sudhan et al.}
\begin{abstract}
Stateful Middleboxes are integral part of enterprise and campus networks that provide essential in-network, security, and value-added services. 
These stateful middleboxes rely on precise network flow identification. However, the adoption of HTTP/3, which uses the QUIC protocol, poses significant challenges to the proper functioning of these devices.
QUIC’s encryption and connection migration features obscure flow semantics, disrupting middlebox visibility and functionality. 
We examine how QUIC disrupts middleboxes like Network Address Translators (NATs), Rate Limiters, Load Balancers, \textit{etc.}, and affects Kubernetes-based service deployments.
To address these challenges, we propose a novel, generalized framework that enables stateful middleboxes to reliably track QUIC connections, even when the endpoints change their internet protocol (IP) address or port numbers.
Our prototype implementation demonstrates that the proposed approach preserves middlebox functionality with HTTP/3 with negligible performance overhead (< 5\%) on both throughput and latency, and works effectively even under high QUIC connection migration rates of up to 100 Hz.

\end{abstract}
\hide{

}
\maketitle
\section{Introduction}
\label{sec:introduction}
The evolution of Hypertext Transfer Protocol (HTTP), particularly the advent of HTTP/3~\cite{rfc9114-http3} 
marks a significant paradigm shift in network communication.
\hide{
Hypertext Transfer Protocol (HTTP) \textemdash{} 
one of the most widely used application-layer protocol
has evolved through various versions, including HTTP/1.1,  HTTP/2 and more recently HTTP/3~\cite{rfc9114-http3}. 
HTTP/1.1 was a significant leap forward from its predecessor, introducing new features like connection reuse with persistent connections, request pipelining, chunked responses, additional cache control \textit{etc.} to improve both performance and efficiency of protocol~\cite{rfc2616}. 
Further, HTTP/2 introduced multiplexing, header compression, and server push to significantly improve resource utilization and speed~\cite{rfc9113}.
However, certain performance issues persist in HTTP/2, including head-of-line blocking and initial handshake latency. Furthermore, HTTP/2 is also susceptible to security vulnerabilities such as continuation frame attack and TLS downgrade attacks~\cite{continuation-attack-h2}.
}
Unlike its predecessors HTTP/1.1~\cite{rfc2616} and HTTP/2~\cite{rfc9113}, which operate over Transmission Control Protocol (TCP), HTTP/3 adopts a new transport-layer protocol, QUIC~\cite{rfc9000-quic}.

\begin{figure}[!t]
    \centering
    \includegraphics[width=0.95\linewidth]{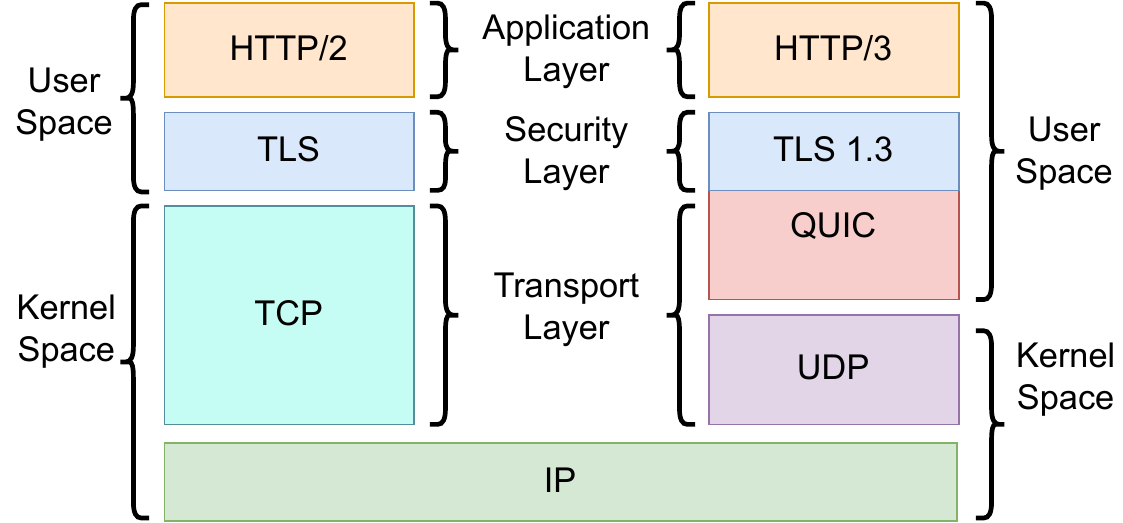}
    \vspace{-1mm}
    \caption{Network Stack of HTTP/2 and HTTP/3}
    \label{fig:http-position}
    \vspace{-5mm}
\end{figure}

Fig.~\ref{fig:http-position} shows the network stack of HTTP/2 and HTTP/3. We can observe that QUIC runs on top of the User Datagram Protocol (UDP) in user space.
This ensures operational compatibility across the diverse existing network appliances that already support UDP and also facilitates ease of deployment~\cite{rfc9000-quic}.
In addition, QUIC incorporates various innovative features such as multiplexing, reduced connection establishment latency, encryption, improved congestion control, loss recovery mechanisms, and seamless migration capabilities. 
These advancements offer significant advantages over TCP, ensuring enhanced security, reliability, and efficiency for HTTP/3, thereby catering to the demands of modern internet applications and improving user experiences across diverse network conditions.
Seamless connection migration is one of the prominent features introduced by QUIC. This facilitates the client to switch from one network address (IP address) to another without the need to alter or reestablish the transport connection. For example, when the client (mobile device) transitions from the campus/ enterprise network to a network provided by a public Internet Service Provider (ISP).
%
Unlike traditional transport protocols, QUIC identifies connections using Connection IDs (CIDs) instead of relying on IP addresses and port numbers. This decoupling enables seamless connection migration in dynamic network environments, such as those involving IP or interface changes.
This transition, while offering enhanced performance, security and seamless connection-migration benefits for end-users, introduces notable challenges for network intermediaries, commonly referred to as middleboxes or network functions.
It was also observed that malicious actors can exploit 
QUIC to orchestrate UDP hole-punching, denial-of-service attacks and also to disrupt network operations~\cite{joarder2024exploring, selvam2024security, chatzoglou2023revisiting}.

Moreover, various critical network functions, known as middleboxes, such as Network Address and Port Translator (NAPT/NAT), Load Balancers (LB), Rate Limiter (RL), among others, depend on connection tracking tables to deliver their intended services. 
These middleboxes internally rely on mapping network flows, which comprise 5-tuple information encompassing source and destination IP addresses, ports, and protocols.
We observe that connection migration—which entails modifying the address and/or port components of the IP 5-tuple—can interfere with the expected behavior of various middleboxes. We demonstrate that such migration enables a QUIC connection to evade flow-based rate limiting. 
Additionally, we illustrate how Layer 4 load balancers may malfunction after a migration event. 
Furthermore, we show that connection migration can trigger denial-of-service (DoS) attacks on standard NAPTs. 
Finally, we also present how this mechanism can disrupt Kubernetes-based deployments, particularly those reliant on consistent network flows.

In response to the critical challenge of mitigating disruptions to middlebox functionality caused by connection migration, this paper proposes a novel framework: QUIC-aware Stateful Middleboxes (QASM). 
The framework enables enhanced tracking and identification of QUIC connections after migration, allowing middleboxes to maintain uninterrupted service delivery in dynamic and heterogeneous network environments. Accordingly, the key contributions of this work are as follows:
\begin{itemize}
\item We identify and analyze the scenarios where QUIC’s seamless connection migration can disrupt the functionality of stateful middleboxes. (\S \ref{sec:vulnerabilities}). 
\item We show that QUIC connection migration can disrupt services and induce denial-of-service (DoS) in stateful middleboxes 
and Kubernetes service meshes. (\S \ref{sec:vulnerabilities}).
\item We propose QASM (QUIC-aware Stateful Middleboxes), a novel framework that enables middleboxes to maintain connection awareness and continuity even in the presence of QUIC connection migration. (\S \ref{architecture}).
\item We implement and evaluate the performance of QASM, and show that QASM introduces negligible overhead while providing correct middlebox operation. (\S \ref{sec:implementation}).
\end{itemize}

\hide{
\section{Background}
\label{sec:background}
In this section, we provide a brief overview on the family of HTTP protocols and QUIC. 
In addition, we briefly describe the operational aspects of a few critical middleboxes that are prevalent in different network environments such as large data centers, campus and enterprise networks.

\subsection{Hypertext Transfer Protocol}
\begin{figure}[ht!]
    \centering
    \vspace{-2mm}
    \includegraphics[height=4.6cm]{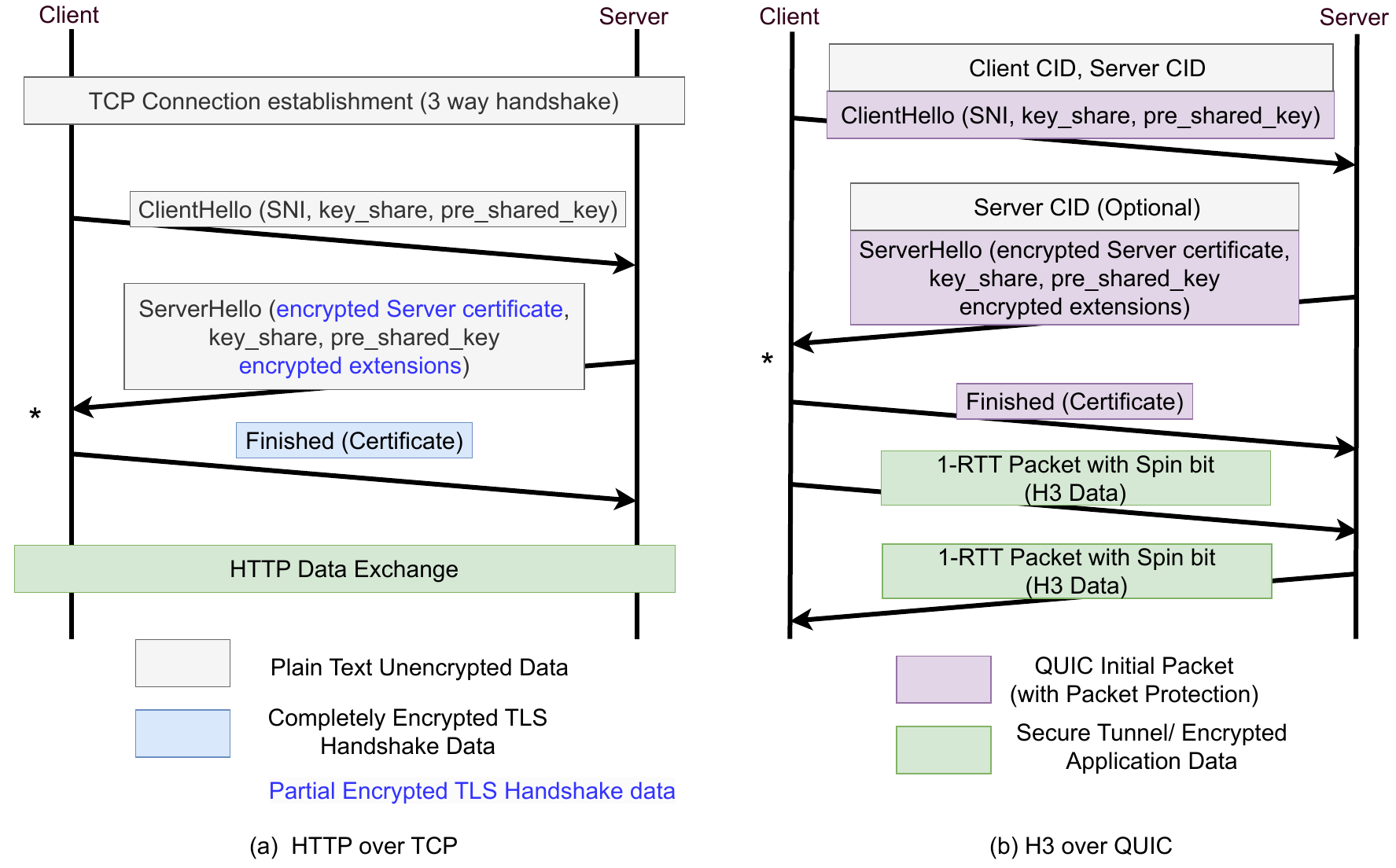}
    \vspace{-2mm}
    \caption{HTTP Protocol handshake when using TCP and QUIC. 
}
    \vspace{-4mm}
    \label{fig: HTTP Flow}
\end{figure}

The HTTP protocol has three prominent versions: HTTP/1.1, HTTP/2, and HTTP/3. 
While HTTP/1.1 and HTTP/2 utilize TCP for transport and leverage TLS for secure connections atop TCP, HTTP/3 employs QUIC as its transport protocol.
Notably, TLS is integrated within the QUIC stack, enabling encryption of critical QUIC headers, with only UDP headers transmitted in unencrypted mode.
Fig. \ref{fig: HTTP Flow} compares the handshake and information flow in HTTP/3 and prior versions of HTTP, namely HTTP 1.1 and HTTP/2.
HTTP/3 only requires 1-RTT entire handshake as QUIC combines the transport and TLS handshakes, significantly reducing connection establishment latency compared to earlier versions of HTTP protocol.

A QUIC connection is identified by a pair of connection IDs (CIDs), which are variable-length fields located in the QUIC header. 
These identifiers operate independently of the IP address and the transport layer port numbers.
This decoupling effectively separates the QUIC connection from its underlying IP address and UDP port numbers.
Unlike TCP, which resets connections on source IP or port changes, QUIC leverages Connection IDs (CIDs) to enable seamless connection migration across network boundaries.
Moreover, QUIC enhances connection privacy by allowing endpoints to use multiple CIDs and recommending the retirement of old identifiers upon network path changes~\cite{rfc9000-quic}.

\subsection{Critical Network Functions (Middleboxes):}
Network infrastructure comprises diverse networking middleboxes that are utilized to enhance performance, security, and user experience. We offer insights into key stateful middleboxes that are pertinent to our research.

\noindent\textbf{Network Address and Port Translation (NAPT/ NAT)}
allows the mapping of local IP addresses to global IPs, enabling private networks to communicate with the Internet using a limited pool of global IP addresses~\cite{rfc1631-nat}. 
These NAPT devices are typically positioned at the edge or egress point of private networks. 
When a new flow (packet transmission from the private network to the internet) is detected, these devices create an entry in the NAT table, mapping local IP addresses and port numbers to global IP addresses and ports.
Dynamic NAT devices, commonly used in large enterprise and campus networks, generate NAT entries dynamically and refresh them periodically, typically after a flow terminates or after a specified period of inactivity.
These entries are refreshed based on a default timeout period of 300 seconds, with a minimum timeout of 120 seconds for UDP connections.~\cite{rfc4787-udp-nat}

\noindent\textbf{Load Balancer (LB)}
is a fundamental networking device widely deployed to evenly distribute network traffic across multiple backend servers to ensure efficient resource utilization and optimal performance. 
LBs operate at either the transport layer (L4) or the application layer (L7, functioning as reverse proxies), and utilize various standard algorithms such as round robin, weighted round robin, least connections,
or least response time to map incoming requests to one of the backend servers~\cite{load-balancer}.
Upon routing a request to a server, the LB maintains a connection record for the session to track the backend server assigned to handle that specific request. 
Periodically, these entries are refreshed to ensure the accuracy and cleanliness of the mapping.

\noindent\textbf{Rate Limiter (RL)}
is a crucial networking mechanism that regulates traffic to enforce predefined limits, helping mitigate congestion, prevent abuse, and maintain quality of service (QoS). 
In flow-based rate limiting, the transmission rate of individual network flows is controlled according to their characteristics~\cite{128t_rate_limiting}. 
This is achieved by inspecting packets at the network layer and applying policies based on IP addresses, ports, and protocol types. 
When flows exceed their configured limits, the rate limiter enforces control through packet drops, transmission delays, or traffic shaping, ensuring fair allocation of bandwidth and preserving overall network stability.

\hide {
\noindent\textbf{Rate Limiter (RL)}
is a crucial networking component designed to control and manage the flow of network traffic, ensuring that it adheres to predefined rate limits. 
It serves to mitigate network congestion, prevent abuse, and enforce quality of service (QoS) policies.
Flow-based rate limiting, a key functionality of a rate limiter, involves regulating the data transmission rate for specific network flows based on their characteristics~\cite{128t_rate_limiting}.
This is typically achieved by inspecting packets at the network layer and applying rate-limiting policies based on source/destination IP addresses, source/destination port numbers, and protocol types.
Upon detecting network flows that exceed the configured rate limits, the rate limiter takes appropriate actions, such as dropping packets, delaying transmission, or shaping traffic, to enforce the desired rate limits. 
This ensures that network resources are fairly allocated among different flows and prevents any individual flow from consuming an excessive amount of bandwidth, thereby maintaining network stability and performance.
}
}

\section{Motivation: Impact of HTTP/3 (QUIC) on Stateful Middleboxes}
\label{sec:vulnerabilities}
\textbf{HTTP/3} the latest version in the HTTP family, makes use of QUIC as its transport protocol.
Notably, QUIC incorporates TLS 1.3 directly into its protocol stack, which allows for end-to-end encryption of essential QUIC headers and payload data. As a result, only the UDP headers are sent unencrypted, which improves security and privacy by keeping connection details hidden from eavesdroppers.
Moreover, HTTP/3 requires only a single 1-RTT handshake to establish a connection because QUIC integrates both the transport-layer and TLS handshakes. 
This consolidation significantly reduces connection setup latency compared to earlier HTTP versions, which required multiple round trips to complete these steps separately.

A QUIC connection is identified by a pair of connection IDs (CIDs), which are variable-length fields located in the QUIC header. 
These identifiers operate independently of the IP address and the transport layer port numbers.
This decoupling effectively separates the QUIC connection from its underlying IP address and UDP port numbers.
Unlike TCP, which resets connections on source IP or port changes, QUIC leverages Connection IDs (CIDs) to enable seamless connection migration across network boundaries.
Moreover, QUIC enhances connection privacy by allowing endpoints to use multiple CIDs and recommending the retirement of old identifiers upon network path changes~\cite{rfc9000-quic}.

In an HTTP/3 connection, stateful middleboxes such as NAPT, LB, RL, \textit{etc.}, face unique challenges in identifying connections due to the use of QUIC CIDs rather than the conventional IP 5-tuple. The QUIC protocol
facilitates the use of multiple CIDs within a single session and permits hosts to dynamically alter their IP addresses or port numbers.
The flexibility offered by QUIC's connection identifiers (CIDs) introduces challenges for middleboxes that depend on connection tracking to function correctly. In contrast to traditional transport protocols, 
where the changes in IP 5-tuple indicate a new connection, the dynamic changes to QUIC's CIDs can lead middleboxes like NAPT or LBs to misinterpret such alterations as distinct (new) connections. Consequently, the operation of these middleboxes, which rely on accurate connection identification, can get significantly compromised.

Moreover, QUIC’s inherent privacy features further exacerbate the challenges for stateful middlebox operations. The use of dynamically assigned and potentially multiple Connection IDs (CIDs) within a single session, coupled with endpoint mobility through IP and port number changes, breaks the traditional 5-tuple-based flow identification. As a result, it undermines the effectiveness of stateful middlebox functions such as NAPT, LB, and RLs, potentially leading to service degradation in QUIC-enabled deployments.

In this section, we first describe the impact of connection migration on various middleboxes, namely, NAPT, RL, LB, and demonstrate how malicious actors can exploit connection migration to circumvent their intended functionality. Furthermore, we examine how connection migration affects Kubernetes-based deployments.

\subsection{Denial of Service on NAT}
\begin{figure}[!htbp]
    \centering
    \includegraphics[width=0.95\linewidth]{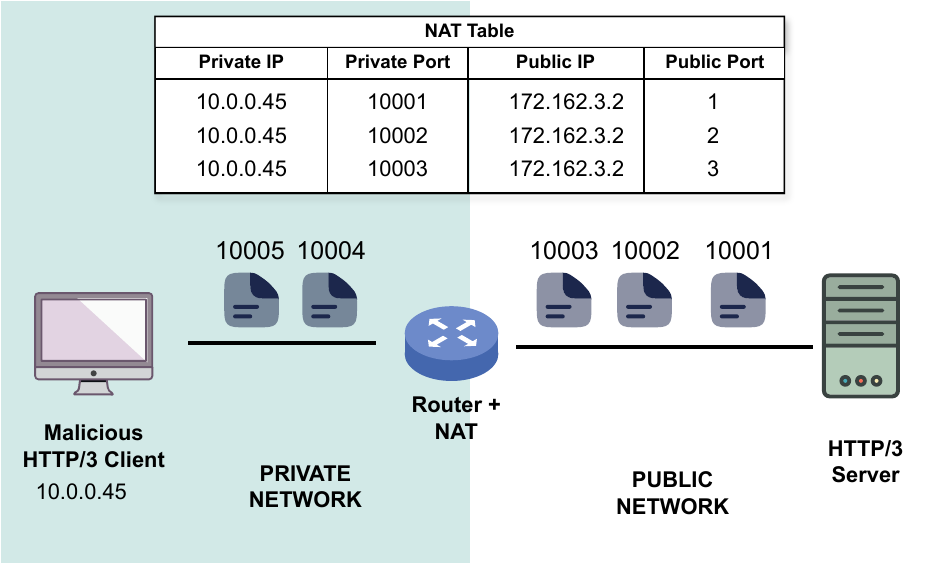}
    \vspace{-2mm}
    \caption{DoS with HTTP/3 and NAT. 
    QUIC packets of same connection with different network address. 
    The number on top of the packet indicates the UDP Source port number.}
    \label{fig:nat-setup}
    \vspace{-4mm}
\end{figure}

Fig.~\ref{fig:nat-setup} illustrates a typical enterprise network configuration, where a private network accesses the Internet through a NAT device.
In such setups, the NAT identifies QUIC connections as standard UDP flows. Upon each connection migration, we noticed that a new entry is inserted into the NAT table, while the original (existing) entry continues to remain active.
However, when the available public IP and port mapping resources are exhausted, the NAT device is no longer able to allocate new mappings for subsequent connections.
This leads to packet drops from the private network, resulting in connection failures and potential DoS conditions.

By exploiting this vulnerability, an attacker can initiate QUIC connections with external servers and begin data transfers. 
By manipulating the underlying UDP connection parameters—specifically, altering the source IP address and UDP port—the client can trigger multiple connection migrations within a short time window. 
Each modification causes the QUIC endpoint to interpret the change as a legitimate migration to a new network path~\cite{rfc9000-quic}.

However, on the source side, the NAT device treats each migration as a new flow, requiring the creation of additional connection entries in its table. 
This necessitates fresh public IP and port mappings for what is logically a single ongoing connection. 
While NAT table resources can often scale dynamically, the primary limitation stems from the finite availability of public IP addresses and port numbers.

Given that the maximum number of available ports is $2^{16}$, and the allocation of public IP addresses to enterprise/campus networks is limited, NATs can only accommodate a fixed number of concurrent mappings. 
Through experimentation using a Mininet emulation of the setup illustrated in Fig. \ref{fig:nat-setup}, we employed an \textit{aioquic}-based HTTP/3 client and server alongside a Scapy-based NAT\footnote{Implementation can be found at \href{https://github.com/hari-19/quic-aware-middlebox}{https://github.com/hari-19/quic-aware-middlebox}}. Our results show that this finite capacity can be easily exhausted within the NAT's timeout period, leading to a DoS scenario that impacts all legitimate users within the network. 

\subsection{Impact on Load Balancers}
\begin{figure}[t]
    \centering
    \includegraphics[width=0.9\linewidth]{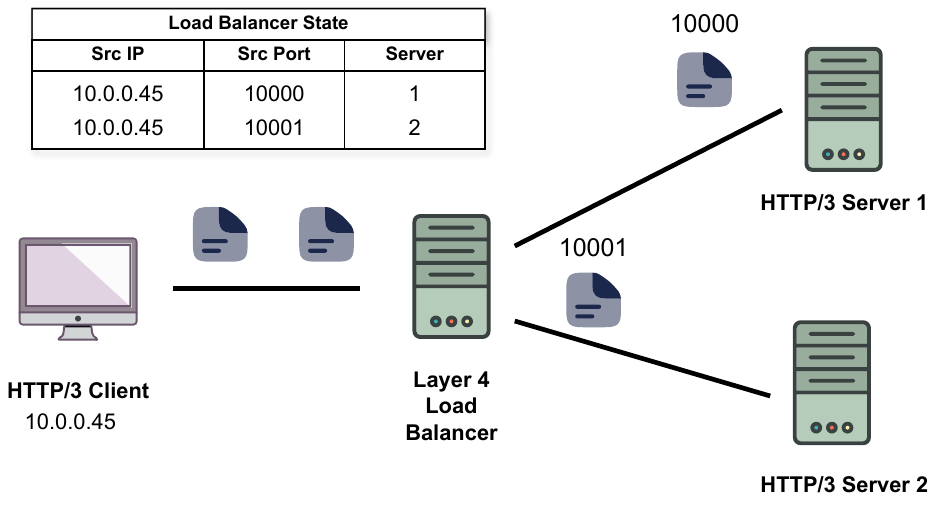}
    \vspace{-2mm}
    \caption{Packets of the same QUIC connection get mapped to different servers post-connection migration. The number on top of the packet indicates the UDP Source port number.}
    \vspace{-5mm}
    \label{fig:lb-setup}
\end{figure}

Many applications on the internet require the persistence of the connections. 
This necessitates the load balancers (LBs) to store the characteristics of the incoming connections, \textit{i.e.}, the IP 5-tuple information for consistent mapping to the backend server nodes. 
In HTTP/3, since a QUIC connection is independent of the traditional connection properties like the IP addresses and port numbers, a single QUIC connection may be identified as a different flow by the LB after a connection migration, resulting in it being mapped to a different backend server, as shown in Fig. \ref{fig:lb-setup}. 
This disrupts the functionality of the LB, as the persistence of the connection can't be maintained. 
To ensure proper operation of LB, it is imperative that a single connection remains mapped to a single server across connection migrations.

In a typical load balancer, backend server selection is often based on a hash of the client's IP address. 
In our experiments using NGINX v1.25.5, configured as shown in Figure~\ref{fig:lb-setup}, the LB was set to distribute incoming UDP flows using IP-hash–based load balancing across two HTTP/3 servers. 
When the connection migration is initiated by the HTTP/3 client, the shift in the source IP and port prompted the load balancer to reroute the traffic to a different backend server rather than the one initially managing the connection. This discrepancy caused the QUIC connection to be terminated, leading to a disruption in service experienced by the user.

Additionally, the tertiary content addressable memory (TCAM) capacity used for connection table is typically limited to a few hundred megabytes~\cite{load-balancer-nuts-bots-f5}.
Due to connection migration, a single QUIC connection may be recorded as multiple entries in a LB's connection table. This introduces a vulnerability, where an attacker can abuse this loophole by repeatedly triggering connection migrations and sending QUIC packets to a server situated behind the LB. Further, this can lead to significant memory usage in the LB (due to new connection tracking entries),  leading to high memory consumption and a resource-exhaustion denial-of-service (DoS) attack. 

\subsection{Impact on Rate Limiters}

\begin{figure}[!htbp]
    \centering
    \includegraphics[width=0.95\linewidth]{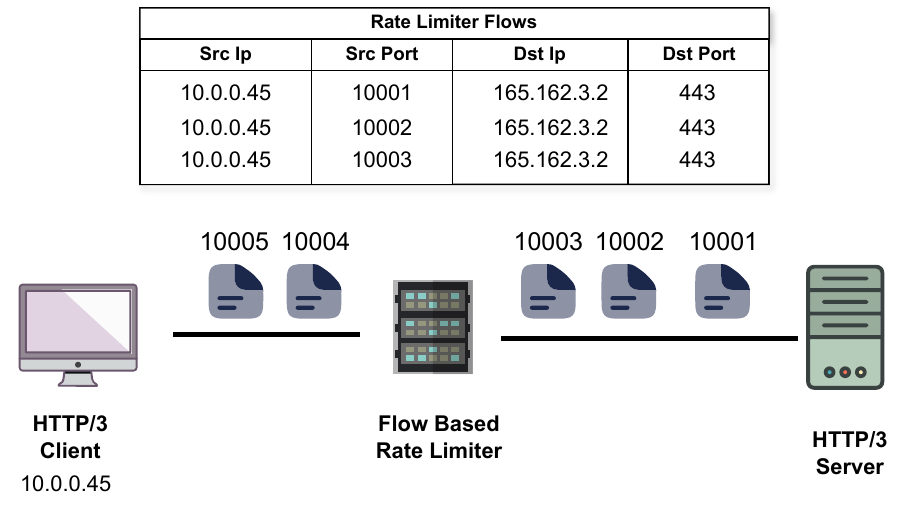}
    \vspace{-2mm}
    \caption{Flow based RL creating multiple flows for a single HTTP/3 connection after connection migration. The number on top of the packet indicates the UDP Source port number.}
    \vspace{-3mm}
    \label{fig:rl-setup}
\end{figure}
Rate limiters (RLs) are one of the critical components within network infrastructures. 
They enforce limits on network traffic to manage congestion, prevent abuse, and uphold quality of service (QoS) standards. 
Flow-based rate limiting, an important feature of RL, involves regulating the data transmission rate for specific network flows based on their unique characteristics. 
Parameters typically governed by network rate limiting include the maximum allowed connections per client, maximum traffic rate (bandwidth) per client, and peak burst size, among others. 
These parameters establish an upper bound on the network traffic that an individual user or connection can generate within a network environment.

Traditionally, network rate limiters (RLs) depend on the IP 5-tuple to maintain connection state and enforce rate limits. However, HTTP/3’s adoption of the QUIC transport protocol, which enables connection migration by allowing changes in client IP addresses and ports without interrupting sessions, fundamentally undermines the reliability of 5-tuple-based identification of connection state, and thereby challenges the effectiveness of conventional rate-limiting mechanisms.
In our experiment with Mininet emulation of scapy-based RL using the token-bucket algorithm, \textit{aioquic} HTTP/3 client and server setup as shown in Fig. \ref{fig:rl-setup}. We identified that when the client migrates to different source IP addresses under the QUIC protocol, the connection tracking information essential for rate limiting becomes disrupted.
As a result, RLs operating outside the application layer struggle to distinguish between migrated QUIC traffic and new connections.
The RL identifies the migrated connection as a new flow instead of the previously existing flow.  
This inability to differentiate between existing and new connections undermines the accuracy of per-connection rate limiting for QUIC traffic, allowing HTTP/3 users to bypass rate limiting.

As a consequence, QUIC connection migration presents significant challenges for RLs in accurately enforcing traffic limits. Without robust mechanisms to accommodate the dynamic nature of QUIC connections, RLs risk ineffectiveness in regulating traffic flow and upholding Quality of Service (QoS) standards in contemporary networks.

\subsection{Impact on Kubernetes-based deployments}
Kubernetes is a widely adopted open-source container orchestration platform that has become a critical backbone for cloud-native applications~\cite{kubernetes}. 
In Kubernetes-based deployments, network connection tracking plays an important role in enabling core network services like NAT, service routing and load balancing.
Each Kubernetes node typically relies on Linux kernel’s conntrack tables to maintain state information for all active network connections. 
This is particularly important for maintaining consistent NAT mappings and routing decisions across services and pods. 
Each node in the cluster maintains a local conntrack table indexed by the IP 5-tuple. However, QUIC's support for connection migration challenges this model. When a QUIC connection undergoes migration, the change in the IP 5-tuple causes the conntrack system to treat it as a new connection, resulting in the creation of a new flow entry.

\begin{figure}[!htbp]
    \centering
    \includegraphics[width=\linewidth]{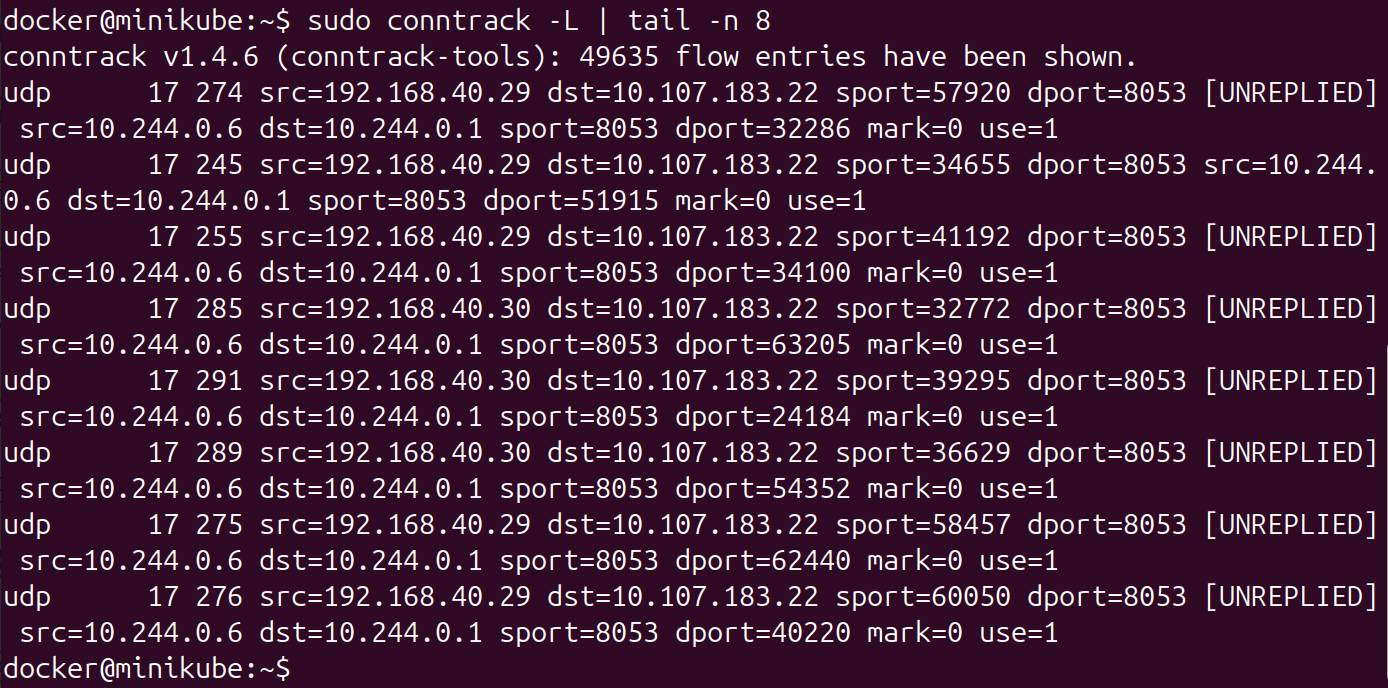}
    \vspace{-3mm}
    \caption{
    Kubernetes Conntrack Flooding by QUIC Migration}
    \label{fig:kub-dos}
    \vspace{-6mm}
\end{figure}

Under high-frequency connection migration scenarios, such as those induced by malicious clients or unstable network conditions, the conntrack table can quickly fill up. 
Fig.~\ref{fig:kub-dos} illustrates how repeated migrations lead to an accumulation of stale or redundant entries.
Once the conntrack table reaches its capacity, the node begins to drop subsequent connection attempts, even from legitimate clients. Thus it severely impacts the service availability and performance across the Kubernetes cluster. 
This vulnerability poses a significant threat in dynamic or large-scale environments where a high volume of short-lived or migrated connections are common, and where the default conntrack table size is limited. 

\section{Architecture and System Design}
\label{architecture}
\label{sec:quic-middlebox}
\subsection{Design Goals}
\label{sec:design-goals}
In designing \name, we focus on the following key desired properties that need to be meet for easier and wider adoption of the proposed solution.

\noindent \textbf{Generality:}
The framework must support a wide range of middleboxes and not be limited to a specific type or configuration. 
Internet infrastructure is inherently heterogeneous, comprising devices such as firewalls, load balancers, traffic shapers, and deep packet inspectors—each with distinct roles and behaviors. 
A narrowly scoped solution would fail in environments with diverse or evolving middlebox deployments. 
By ensuring generality, the framework remains compatible with both existing and future middleboxes, enabling broader adoption of QUIC without requiring constant updates or custom adaptations.
    
\noindent \textbf{Efficiency:}
The framework must incur no or minimal overhead on normal processing.
The addition of this framework should not have any perceivable impact on the performance of the end hosts.
Minimizing the overheads is critical for user adoption. 
If users experience performance degradation after enabling the framework, they're less likely to continue using it. 
This could hinder the overall adoption of the framework, even if it mitigates the network service disruptions.

\noindent \textbf{Lightweight:}
The framework should be lightweight, meaning it should introduce minimal overhead on the packet processing of middlebox  compared to default configurations. 
This is crucial because any increase in processing time per packet can significantly impact the middlebox's performance.
Higher overhead translates to increased latency, reduced throughput of the middlebox. 
If the framework adds higher processing cost, the middlebox might not be able to keep up with incoming traffic, leading to dropped packets and bottlenecks.

\noindent \textbf{Scalability:}
In a typical network, the number of QUIC flows is expected to be large and constantly growing. 
This, coupled with an increasing number of middleboxes, can lead to performance degradation if the framework isn't scalable.
Scalability ensures efficient distribution of processing load, minimizing impact on performance and maintaining smooth user experience even under high traffic.

\noindent \textbf{Transparency:}
The clients should be agnostic to the existence of the middleboxes. 
Clients can focus on their core tasks without needing to worry about the intricacies of network infrastructure. 
This promotes cleaner and more maintainable code.
By not relying on the specific functionalities of middleboxes, clients can work across different network architectures, ensuring that they function seamlessly regardless of existence or change in the middlebox.
If a client relies on a specific middlebox functionality and that middlebox malfunctions, communication can break down. Agnostic clients avoid this single point of failure, making communication more reliable.
    
\noindent \textbf{Non-invasiveness:}
The framework should be deployable without requiring any modifications to the QUIC protocol library itself. 
This non-invasive approach is crucial for several reasons. QUIC has several implementations, such as \textit{msquic}, \textit{aioquic,} and \textit{quic-go}. 
Requiring changes to the QUIC library would necessitate updates to each of these codebases, leading to significant delays in adoption.
A non-invasive framework simplifies deployment in existing network infrastructures. 
There's no need to update or modify QUIC implementations on end devices or servers, which streamlines the integration process and reduces potential compatibility issues.
    


\subsection{Tracking Attributes}
QUIC packets include a Destination Connection ID (DCID), which identifies the connection from the receiver’s perspective.
We refer to the DCID generated by the client in its initial packet during connection establishment as the Original Destination Connection ID (O-DCID). 
This ID is unique to the client’s view of the connection and is used to connect to a specific server. 
As such, the combination of the O-DCID and the client’s address (IP and port) uniquely identifies a QUIC connection from the client's side.
Throughout the lifetime of a QUIC connection, the client may change its DCID as part of the connection migration process. To maintain continuity, we associate all DCIDs used by the client with the original tuple comprising the \textbf{O-DCID} and the client's address.


On the server side, DCIDs are generated independently by each server and are not guaranteed to be globally unique. However, the combination of a DCID and the server’s network address (IP and port) is guaranteed to be unique. This is because the server’s address remains constant for the duration of the connection, and the DCID uniquely identifies the connection on that specific server.
Therefore, to reliably track a QUIC connection across migrations and address changes, the framework must maintain the following attributes for each connection: i) O-DCID, ii) Set of DCIDs, iii) Set of client addresses, iv) Server IP and port.

\subsection{System Architecture}

\begin{figure*}[!htbp]
\centering
\includegraphics[width=0.85\linewidth]{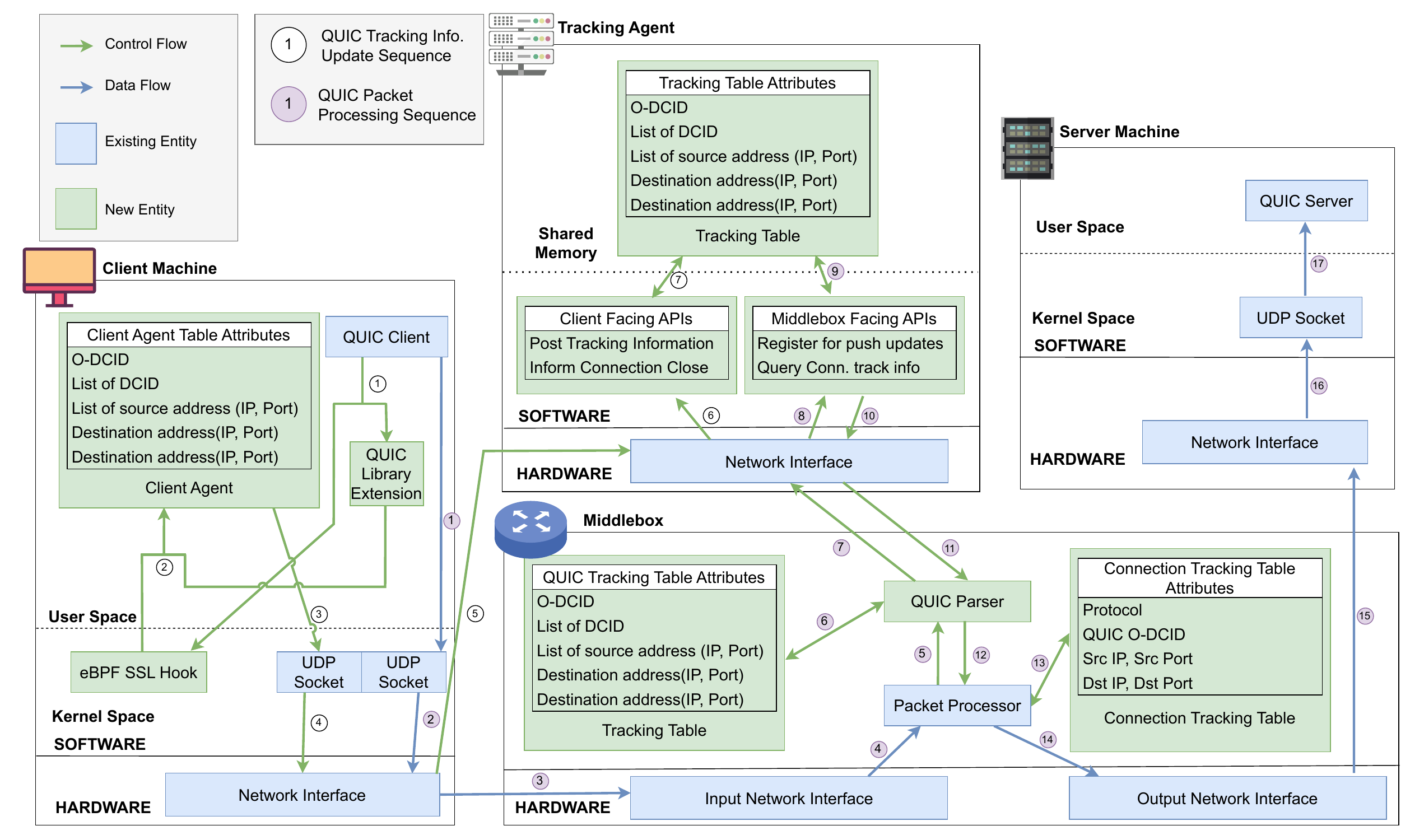}
\vspace{-2mm}
\caption{
Architecture and interaction of components in the proposed framework consisting of i) QUIC Client, ii) Client agent, iii) Tracking agent, iv) Middlebox, and v) QUIC server. 
}
\vspace{-4mm}
\label{fig:quic-middlebox}
\end{figure*}

The architecture of our proposed framework that enables QUIC-awareness in middleboxes comprises three core components: the Client Agent, the Tracking Agent, and the QUIC-aware stateful middlebox, as shown in Fig. \ref{fig:quic-middlebox}. 
The Client Agent collects connection tracking information from the client system and communicates it to the Tracking agent. 
The tracking agent maintains up-to-date connection tracking information from all clients in the system and communicates to the middlebox, which can then utilize it to function properly.

\subsubsection{Client Agent}
The Client Agent is an integral part of the client system. It is responsible for monitoring DCID changes across all QUIC connections initiated by the client.
Its primary role is to facilitate the distribution of the tracking information to the tracking agent.
As the client receives or switches DCID during the lifetime of the connection, the Client Agent asynchronously transmits the updated tracking information to a preconfigured tracking agent in the network.
This ensures that real-time updates regarding changes in DCID are relayed to the tracking agent for effective monitoring and management.
We propose the Client Agent implementation and deployment options as discussed below:

\textbf{\textit{As a part of QUIC Library:}}
The Client Agent can be implemented as a part of the user-space QUIC libraries that are dynamically linked in the user system.
Any QUIC client is expected to use one of the many popular implementations of QUIC available like \textit{msquic, aioquic, quic-go, ngtcp2}, \textit{etc.}
When the client agent is implemented as part of these QUIC libraries, any client program that makes use of these libraries will allow the client agent to retrieve and communicate the tracking information effectively to the tracking agent.

\textbf{\textit{As eBPF kernel module:}}
The Client Agent can be implemented as an extended Berkley Packet Filter (eBPF)~\cite{gbadamosi2024ebpf} kernel module that is independent of specific QUIC implementations.
By using user-space instrumentation techniques such as uprobes to hook into QUIC libraries at points where packets are decrypted (e.g., calls to BoringSSL's EVP\_AEAD\_CTX\_open),
the decrypted QUIC packets can then be inspected to extract NEW\_CONNECTION\_ID frames, which carry the DCIDs issued by the server.
The agent can map these identifiers to individual QUIC connections and relay them to the Tracking Agent.
This user-space tracing approach requires minimal modification to QUIC libraries and can be deployed non-invasively across popular Linux distributions such as Ubuntu, Red Hat, and Manjaro.


Regardless of the implementation approach, the Client Agent operates transparently within the client system, ensuring seamless functionality without impacting the client application program.
By facilitating the efficient distribution of tracking information, the Client Agent plays a crucial role in enabling robust connection monitoring and management within QUIC-enabled environments. 
Its presence ensures timely updates and accurate tracking of connection ID changes.

\subsubsection{Tracking Agent}
The Tracking Agent is a central component of the framework, responsible for maintaining real-time tracking information for active QUIC connections. 
Positioned externally to the client system, its primary role is to maintain an up-to-date mapping of connection metadata and to efficiently relay this information to relevant network middleboxes in a timely manner.
By introducing a dedicated Tracking Agent, we decouple middlebox interaction from the client.
This design choice ensures that client systems remain agnostic to the presence or behavior of middleboxes, preserving end-host transparency and simplifying deployment.
To support scalability, the Tracking Agent does not need to be a monolithic component. It can be sharded based on DCID, enabling distributed processing and parallelism across large-scale deployments.
The Tracking Agent must expose two classes of APIs—Client Agent–Facing APIs and Middlebox–Facing APIs—to facilitate communication with client agents and middleboxes, respectively.

\textbf{\textit{Client Agent–Facing APIs:}}
The client agent communicates with the Tracking Agent using two APIs. 
The first API allows the client to post tracking information, including updates to the DCID or client address during the lifetime of a QUIC connection. 
The second API notifies the Tracking Agent when a connection is closed. Upon receiving this data, the Tracking Agent stores it in a data structure optimized for efficient and low-latency lookups.

\textbf{\textit{Middlebox–Facing APIs:}} 
To support middlebox functionality, the Tracking Agent provides two APIs. 
The first is a query API that accepts a 5-tuple—comprising the DCID, source IP and port, and destination IP and port—to retrieve the latest tracking information for a QUIC flow. 
The second is a subscription API, which enables middleboxes to register for asynchronous notifications about updates to specific connections, allowing proactive reconfiguration in response to connection migration events.

In the standard enterprise network environment, configuring the Tracking Agent to coexist with the DHCP server or the gateway offers several deployment advantages. 
This strategic placement streamlines deployment processes and facilitates seamless integration with existing network infrastructure. 
The proximity to the DHCP server or gateway ensures efficient communication and coordination with network devices, optimizing overall network performance and management efficiency.
By configuring the Tracking Agent to coexist in the gateway, the solution can be extended to a WAN setup where any middlebox in the WAN can query the agent using the public IP of the gateway.

\subsubsection{QUIC-aware Stateful Middlebox}
The QUIC-aware Stateful middleboxes (QASM) retrieve the connection tracking information, specifically the O-DCID, List of Src Address (IP and Port), List of DCID, Dst IP, and Dst Port, from the Tracking Agent, and use this information to perform its operation.
The middleboxes must have a QUIC parser that can retrieve the DCID from the packet.
It must also maintain a tracking table that keeps track of the QUIC tracking information.
The middlebox, thereby, can leverage this tracking table to retrieve O-DCID and perform connection tracking efficiently.

Since the DCID is a variable-length field in QUIC short header packets, implementations might require the DCID length to be communicated from the client agent via the tracking agent.
A lookup of the DCID length based on the source IP and source port can be used to retrieve the DCID from QUIC short header packets.
Once the middlebox extracts the DCID from a QUIC packet and retrieves the corresponding (O-DCID, Src IP, Src Port) from the Tracking Agent, this information uniquely identifies a single QUIC connection.
Along with the IP 5-tuple, QASM leverages the O-DCID, which serves as the QUIC connection identifier, to support efficient and reliable connection tracking in middleboxes.


In scenarios involving a return flow from the QUIC server, middleboxes can also leverage the stored client IP and port number from the forward flow to retrieve the connection mapping. 
In case tracking information is not available for any flow with the tracking agent, the middlebox can consider this a new flow that has not been updated with the tracking agent.
This allows the QUIC connection to be unaffected in case of a packet loss between the client and the tracking agent.
This comprehensive approach ensures efficient and accurate connection tracking within the QUIC environment, enabling seamless operation of middleboxes while facilitating robust network management and security measures.

Table \ref{tab: new-nat} provides an example of a NAT table configuration that would be produced on a QUIC-aware NAT device, showcasing its ability to allocate a single (Public IP, Public Port) pairing for multiple (Private IP, Private Port) combinations associated with a singular QUIC connection. 
This innovative approach optimizes resource utilization in NAT environments by efficiently managing QUIC connections, minimizing address and port allocation overhead. The QUIC-aware NAT's allocation strategy enhances performance and reliability by reducing address exhaustion and port conflicts.

\begin{table}[htbp!]
\caption{Connection Tracking table in a QUIC-aware NAT}
\vspace{-2mm}
\centering
\begin{tabular}{|p{1cm}|p{1cm}|p{1cm}|p{1cm}|p{1.3cm}|p{1cm}|}
\hline
\textbf{QUIC O-DCID} & \textbf{L4 Protocol} & \textbf{Private IP} & \textbf{Private Port} & \textbf{ Public IP} & \textbf{Public Port} 
\\ \hline
\multirow{3}{*}{fa12ab} & \multirow{3}{*}{QUIC} & 10.0.0.45 & 10001 & \multirow{3}{*}{65.12.81.14} & \multirow{3}{*}{19450}
\\ \cline{3-4}
& & 10.0.0.45 & 10002 & &
\\ \cline{3-4}
& & 10.0.0.45 & 10003 & & 
\\ \cline{3-4}
& & 10.0.0.46 & 10000 & & 
\\ \hline
NULL & TCP & 10.0.0.21 & 15000 & 65.12.81.14 & 42005
\\ \hline
NULL & UDP & 10.0.0.25 & 14005 & 65.12.81.14 & 32006
\\ \hline
\end{tabular}
\label{tab: new-nat}
\vspace{-4mm}
\end{table}

\section{Implementation}
\label{sec:implementation}
We implemented scapy-based default and QUIC-aware middleboxes for NAT and RL on a Mininet~\cite{kaur2014mininet} emulated network.
We implemented the Client Agent as part of the \textit{aioquic} library~\cite{aioquic} and the Tracking Agent as a standalone application running on a Mininet host machine. 
An HTTP/3 client-server \textit{aioquic} application is used to evaluate the middleboxes. 
We implemented and evaluated the following two approaches of the proposed framework. (For convenience, we have listed the associated pseudocode in \S \ref{sec:pseudocode}).

\textbf{\textit{Reactive Mode:}}
Whenever the middlebox receives a QUIC packet with unknown tracking information, it queries the tracking agent for the tracking information. The middlebox stores this information locally for faster lookup.    
Whenever the client migrates its connection, the Client Agent sends the connection tracking information to the Tracking Agent. 
When the middlebox receives a QUIC packet with an unknown O-DCID, it queries the Tracking Agent for the (DCID, O-DCID).
Once the middleboxes retrieve the connection tracking information, it can store it locally for faster lookup.
This approach allows the middleboxes to optimize for memory efficiency, allowing them to store tracking information on demand. 

\textbf{\textit{Proactive Mode:}}
When the middlebox receives a packet with unknown tracking information, it queries and registers this information for any future updates of that connection.
The tracking agent would then asynchronously push all new updates of that QUIC connection to the middlebox, making the tracking information available at the middlebox prior to packet processing.
The middleboxes register with the Tracking Agent for push updates.
Whenever the Client Agent updates the connection tracking information to the Tracking Agent, the Tracking Agent pushes this information asynchronously to all the registered middleboxes, making the connection tracking information available to the middleboxes.
This allows the middleboxes to encounter less overhead during packet processing as the connection tracking information will be available with the middleboxes prior to arrival and processing of the packet.

\section{Evaluation}
\label{sec:evaluation}
In this section, we present the results of the evaluation. We conducted to assess the effectiveness of our proposed framework when compared to default middlebox configurations. 
Our objective was to analyze how our proposed framework performs in comparison to traditional middlebox configurations. 

\subsection{Correctness}

The proposed framework QASM presents a robust solution for accurately tracking QUIC flows across multiple connection migrations to ensure precise and consistent flow identification capabilities within the stateful middleboxes.
Our evaluation of the QUIC-aware NAT revealed that each QUIC connection consistently retained a singular (public IP, public port) association throughout multiple connection migrations. 
Despite these migrations, the QUIC client-server application maintained seamless connectivity and successfully facilitated data exchange operations.

\begin{figure}[!htbp]
    \centering
        \subfigure[Default RL and QUIC-aware RL (5 packets/sec/connection).]{
        \includegraphics[height=3.5cm]{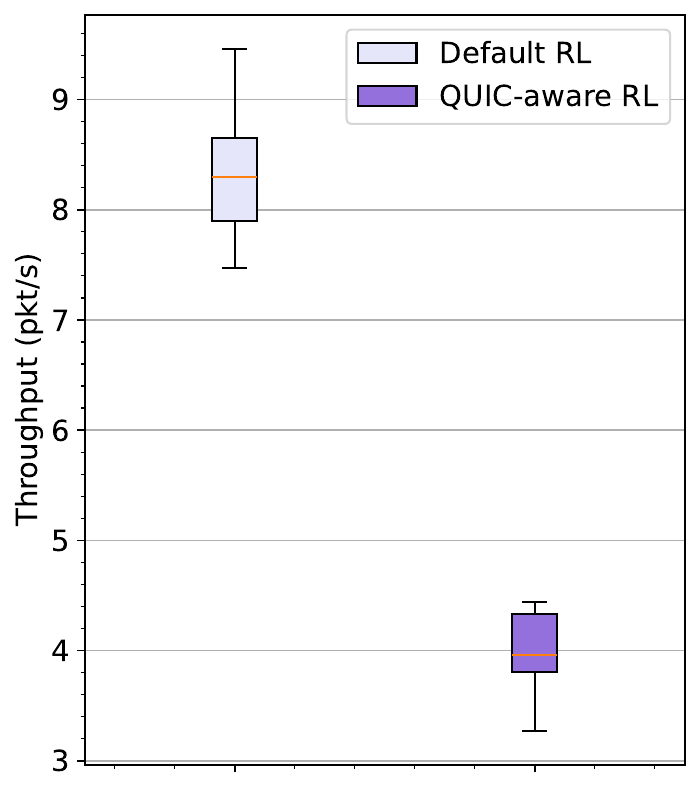}
        \label{fig:throughput-rl}
    }
    \hfill
    \subfigure[Throughput with Default NAT and QUIC-aware NAT.]{
        \includegraphics[height=3.5cm]{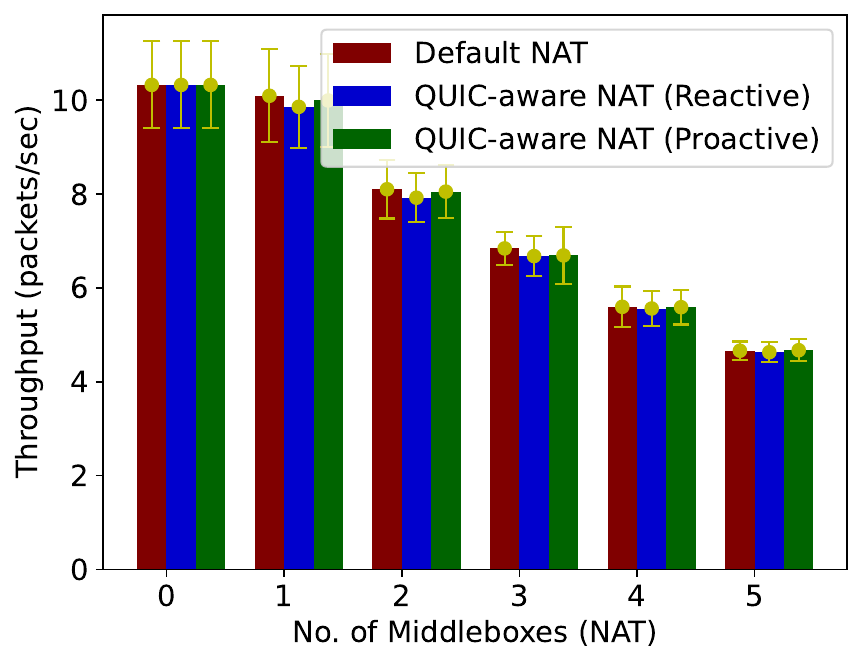}
        \label{fig:throughput-nat}
    }
    \vspace{-2mm}
    \caption{QUIC Throughput with different NAT and RL configurations.}
    \label{fig:combined-throughput}
    \vspace{-6mm}
\end{figure}

Similarly, within our QUIC-aware RL, we observed the unique identification of QUIC connections across connection migrations. 
The flow-based rate-limiting mechanisms were effectively applied to these connections, preventing any attempt to bypass the rate limiter through connection migration.
We configured the rate limiter to have a rate limit of 5 packets per second per connection.
Our QUIC client transfers data while performing connection migrations.
As shown in Fig. \ref{fig:throughput-rl}, in the case of default RL, the QUIC connection was able to bypass the rate-limiting by change of IP 5-tuple, whereas, in the case of QUIC-aware RL, the rate-limiting was honored by the QUIC connections.

These empirical findings underscore the efficacy of our framework in managing QUIC connections within diverse network infrastructures, affirming its practicality for real-world deployment scenarios.


\subsection{Performance}
\subsubsection{Latency}


We measured the latency experienced by the QUIC packets in the middleboxes with different operational settings: Default NAT, QUIC-aware NAT in reactive mode, and QUIC-aware NAT in proactive mode.
In all cases, the client sends 1000 QUIC packets and performs connection migration after every 10 QUIC packets.

\begin{figure}[t!]
    \centering
        \subfigure[Latency in scapy Default NAT and QUIC-aware NAT.]
    {
        \includegraphics[height=4cm]{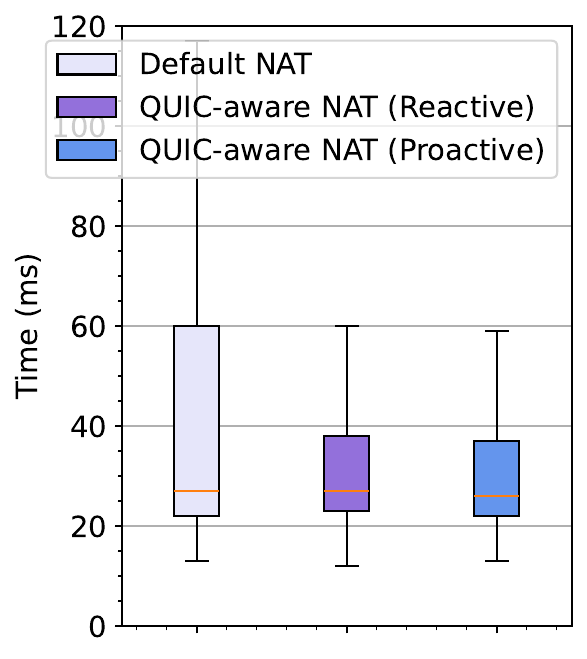}
        \label{fig:nat-latency}
    }
    \hfill
    \subfigure[Latency in scapy Default RL and QUIC-aware RL.]
    {
        \includegraphics[height=4cm]{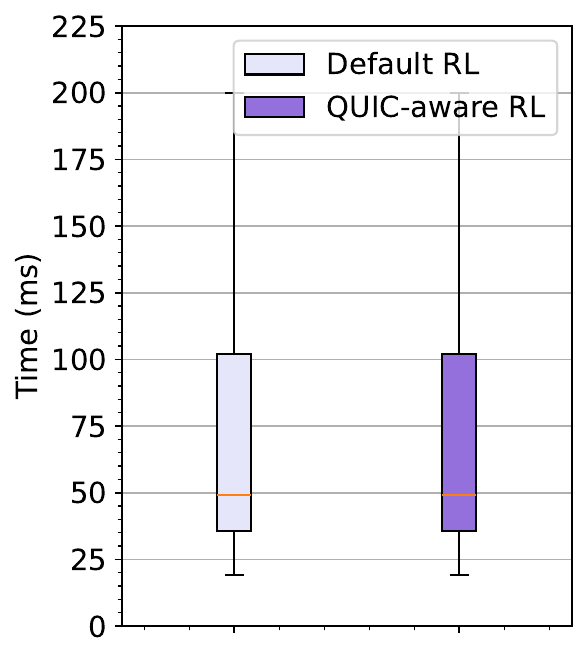}
        \label{fig:rl-latency}
    }
    \vspace{-2mm}
    \caption{Per packet Latency QUIC in different NAT and RL configurations.}
    \label{fig:combined-latency}
\end{figure}

From Fig. \ref{fig:nat-latency}, we can observe that the average latency experienced in the proposed QUIC-aware NAT in both reactive and proactive modes is almost similar to the latency experienced in default NAT.
Additionally, during connection migration, the default NAT exhibits a modest increase in latency relative to the QUIC-aware NAT, also resulting in greater variability in its performance.
Although this appears counter-intuitive, it is explained by the fact that, in the default NAT scenario, the continuously rising number of entries in the NAT table contributes to higher latency. Conversely, in QUIC-aware NAT, where the number of entries remains constant, latency is unaffected.
In typical connections where the connection migration is low, the latency experienced by both the Default NAT and the QUIC-aware NAT are similar as the packet processing for lookup in both cases is similar. 
Therefore, 
the incorporation of QUIC awareness within middleboxes results in a negligible effect on latency.
%

In the case of RL, we measured the latency experienced by the QUIC packets in Default and QUIC-aware configurations.
Just like in the NAT, the client performed connection migration every 10 packets.
From Fig. \ref{fig:rl-latency}, we can observe that the latency is unaffected by the QUIC awareness of the middlebox.

%
%
%
\subsubsection{Throughput}

We measured the throughput in networks consisting up to five NAT middleboxes between the QUIC client and the server.
As base case, we had a network with a simple scapy packet forwarder (0 NAT). 
The client program migrates its QUIC connection every 10 seconds. 
Fig. \ref{fig:throughput-nat} shows the throughput experienced by the Default NAT and the QUIC-aware NAT in reactive and proactive configurations.
From the experiment, it is observed that even in cases of extreme connection migration, the throughput of the QUIC-aware NAT is not affected significantly in comparison with the default NAT.
In typical networks, where the connection migrations are expected to be lower, and the number of middleboxes is greater than one, the QUIC-aware middleboxes can perform as well as default middleboxes.


\subsection{Overhead Analysis: Packet Processing Time}
\label{sec: packet processing overhead quic-aware}

\begin{figure}[!htbp]
    \centering
    \subfigure[Time to process QUIC packets across all cases. \label{fig:proc_total_nat}]{
    \includegraphics[height=3.5cm]{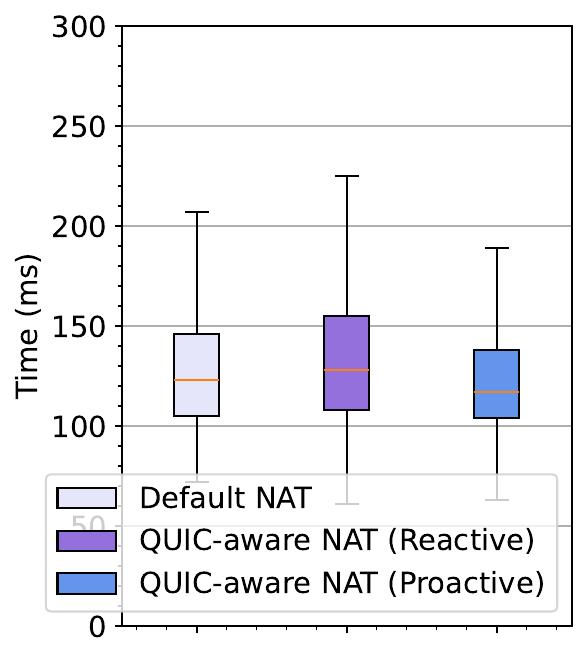}
    }
    \hfill
    \subfigure[CDF of total time to process QUIC packets across all cases\label{fig:proc_cdf_nat}]{
    \includegraphics[height=3.5cm]{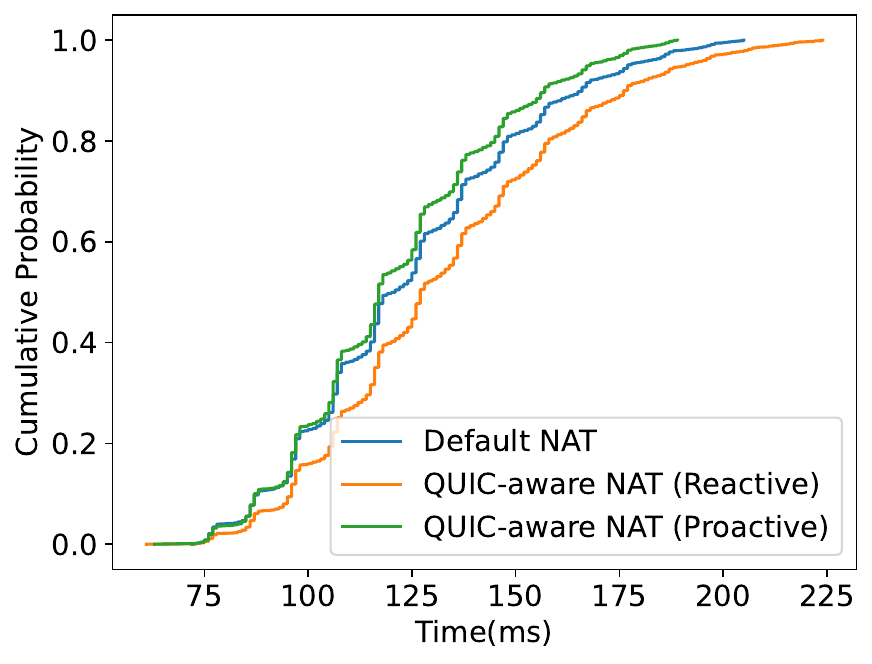}
    }
    \vspace{-2mm}
    \subfigure[Breakdown of the processing Time for QUIC packets in each case. \label{fig:proc_case_nat}]{
    \includegraphics[height=3.1cm]{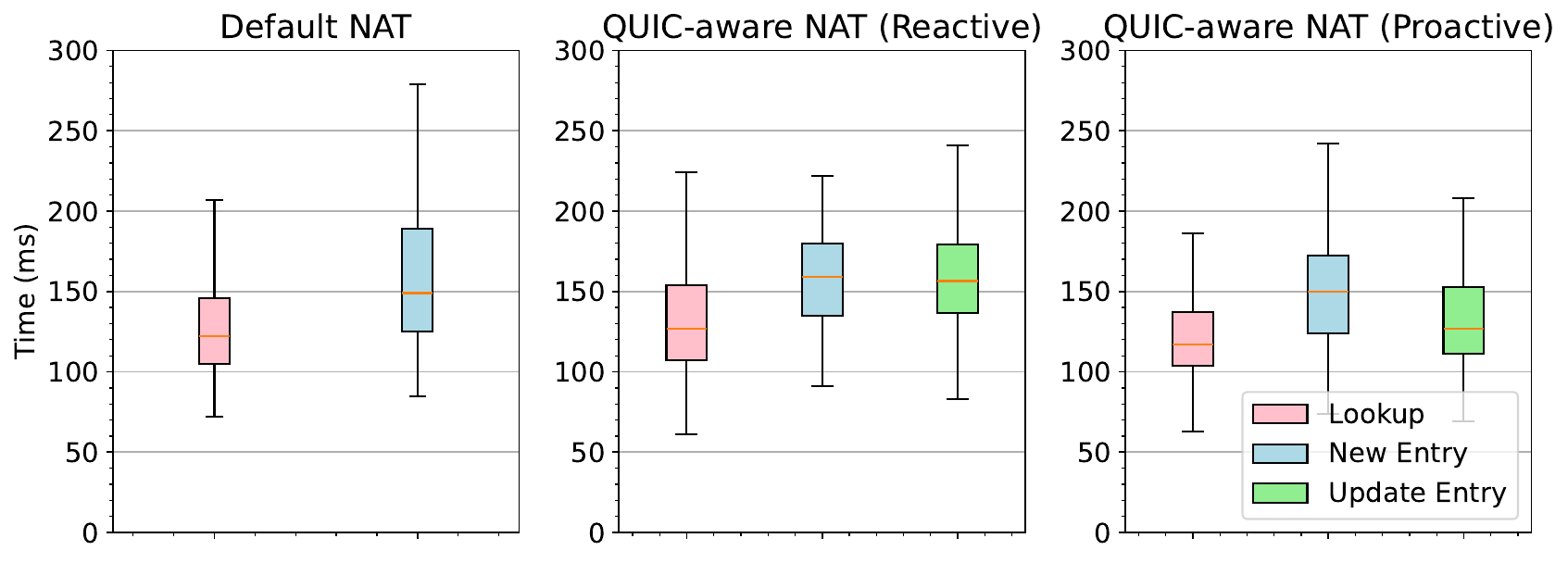}
    }
    \vspace{-2mm}
    \caption{
        Time taken to process QUIC packet in scapy for default NAT and the QUIC-aware NAT (1GHz processor).
    }
	\label{fig:naive-vs-quic-proc}
    \vspace{-6mm}
\end{figure}

We analyze the overheads incurred for packet processing in the NAT middlebox with our two proposed approaches.
We measure the time taken by NAT to process and forward one packet with and without our proposed approach.
Fig. \ref{fig:naive-vs-quic-proc} illustrates the time (in ms)  observed in our implementation of the default NAT, QUIC-aware NAT in reactive mode and QUIC-aware NAT in proactive mode for processing QUIC packets. 
Our results indicate that the QUIC-aware NAT in reactive mode consumes approximately 5\% more time compared to the default NAT, while QUIC-aware NAT in proactive mode performs similar (<1\% overhead) to the Default NAT.

From Figure \ref{fig:proc_case_nat}, we can observe that the time for lookup in the default NAT and the QUIC-aware NAT are approximately the same. 
This indicates that in typical cases, the QUIC-aware NAT doesn't add any additional overhead to the default NAT.
However, the time for creating a new entry in QUIC-aware NAT in a reactive mode is slightly higher than the default NAT, while the QUIC-aware NAT in proactive mode does not add any overhead.

Further, the time for updating an existing entry (updating the private IP and port for an existing entry) is significantly lower in the QUIC-aware NAT in proactive mode than in the QUIC-aware NAT in reactive mode.
We attribute this reduction to be caused by the prior availability of the (DCID, O-DCID) in the NAT memory.

Interestingly, Fig. \ref{fig:proc_total_nat} proves the QUIC-aware NAT implementation demonstrates that the process of retrieving the DCID from QUIC headers and accessing the O-DCID from the tracking agent does not significantly increase complexity. 
While the performance evaluation conducted with the scapy NAT may not directly reflect real-world scenarios, it does highlight that integrating QUIC connection awareness into middleboxes does not introduce substantial additional complexity. 
Consequently, our framework holds promise for practical deployment in real networks with minimal impact on performance.

\section{Related Works}
\label{sec:related-works}
QUIC-LB, currently an IETF internet-draft~\cite{quic-lb}, addresses load balancer service disruptions by embedding routing information within structured connection IDs (CIDs). 
It involves three entities: the server, the load balancer, and a configuration agent, which distributes server ID allocation, routing algorithms, and optional encryption keys. 
Without encryption, intermediate entities can extract server IDs from CIDs, raising privacy concerns. 
Load balancers decode this information to route packets to the correct server. While this approach is effective for load balancing, 
it lacks support for per-flow identification and is thus unsuitable for integration with other stateful middleboxes such as NATs, rate limiters, and intrusion detection systems (IDS).
As a result, a more generalized framework is needed to support connection tracking across diverse stateful middleboxes. 
In~\cite{kistenmacher2025quic}, the authors present two novel attacks that compromise the privacy guarantees of QUIC. The first reveals the number of backend servers behind QUIC-aware load balancers, while the second breaks CID unlinkability, enabling user tracking across network boundaries. 
Their analysis shows that a majority of observed load balancers are vulnerable to these attacks. 

In \cite{selvam2024security} 
the authors highlights that QUIC’s use of Connection IDs (CIDs) can disrupt stateful middleboxes and also 
proposed a partial solution for NAPT devices, suggesting the use of DCID to track connections instead of the traditional IP 5-tuple.  
However, this approach fails when the DCID changes during connection migration \textendash{} a feature permitted by QUIC for privacy and mobility. Consequently, the solution cannot maintain state continuity across migrations. 
Thus, a robust mechanism that tracks the connections after a DCID change is required.
In \cite{ma-sarman-thesis}, the authors propose a stateful firewall that leverages QUIC's path challenge mechanism to manage connection entries in the state table. This design prevents the table from being flooded by regular QUIC packets. However, it remains susceptible to targeted attacks wherein an adversary could flood the state table using carefully crafted path challenge packets. Additionally, the firewall drops other packets during the path validation phase, introducing further latency after connection migration.


\section{Conclusion}
\label{sec:conclusion}
In this work, we identified and demonstrated how QUIC’s connection migration feature can disrupt the functionality of various stateful middleboxes and Kubernetes-based deployments. To mitigate this, we proposed QASM, a generalized framework that leverages QUIC Connection IDs to enable accurate flow tracking across endpoint migrations to ensure correct and consistent middlebox behavior.
QASM introduces a migration-aware tracking mechanism that allows middleboxes to maintain connection state despite IP or port changes.
Performance evaluations show that QASM incurs negligible overhead providing the latency and throughput characteristics comparable to default behavior. 
Our results demonstrate that QASM offers an effective and lightweight solution for maintaining robust middlebox functionality in the face of evolving transport-layer protocols like QUIC.

\bibliographystyle{ACM-Reference-Format}
\bibliography{Bibiliography}

\appendix

\section{Implementation Details}
\label{sec:pseudocode}
The middleboxes for NAT and RL were implemented using Scapy on a Mininet~\cite{kaur2014mininet} emulated network.
The Client Agent was integrated into the \textit{aioquic} library~\cite{aioquic}, while the Tracking Agent runs as a standalone process on a Mininet host.
An HTTP/3 client-server application built with \textit{aioquic} was used for evaluation.

Two approaches were employed for middlebox tracking:

\textbf{\textit{Reactive Mode:}}
The middlebox queries the Tracking Agent only when a QUIC packet arrives with unknown tracking information. Retrieved mappings are cached locally for subsequent packets. This mode prioritizes memory efficiency and on-demand updates. Pseudocode detailing this behavior is provided in Algorithm~\ref{algo:middlebox_api}.

\textbf{\textit{Proactive Mode:}}
Upon encountering a QUIC packet with unknown tracking information, the middlebox registers for updates for that flow. The Tracking Agent asynchronously pushes all new updates of the corresponding QUIC connection, ensuring the middlebox has the tracking information prior to packet processing. Algorithm~\ref{algo:middlebox_api_offline} summarizes the procedure.

\subsection{Client Agent}

We extended the \textit{aioquic} v1.0.0 library to hook into the API that changes the connection ID to implement our Client Agent. 
In \textit{aioquic}, the client can change the peer's connection ID (DCID) using the API, \texttt{change\_connection\_id()}.
This API in turn uses \texttt{consume\_peer\_cid()} function to retire the existing CID and switches the connection to use the next available CID.
Our Client Agent attaches to this hook, extracting the new peer CID for the connection and communicating it to the Tracking Agent as given in the Algorithm \ref{algo:client_agent}.

\begin{algorithm}[!ht]
	\caption{Client Agent}
\textbf{Input:} dcid, o-dcid, ip\_tuple extracted from quic client
	\begin{algorithmic}[1]
    \State agent\_ip, agent\_port = \textit{pre-defined in Environment}
    \State tracking\_data = pack(dcid, o-dcid, ip\_tuple) 
    \State socket = Socket()
    \State socket.send(tracking\_data, to = (agent\_ip, agent\_port))
    \State socket.close()
	\end{algorithmic}
     \label{algo:client_agent}
\end{algorithm}

\subsection{Tracking Agent}
The Tracking Agent is implemented as a standalone Python program running on a host machine. It exposes two distinct APIs:

Client-Facing API: Receives updates from the Client Agent regarding changes in connection identifiers.

Middlebox-Facing API: Responds to queries from middleboxes or proactively pushes updates for registered connections.

Both APIs run on separate threads—namely, the Client thread and the Middlebox thread—to allow concurrent processing of requests. This multithreaded design ensures high responsiveness under heavy request loads. The mappings between DCID and O-DCID are stored in a shared Python dictionary, accessible by both threads.

\paragraph{Client-Facing API}
The Client-Facing API receives messages from the Client Agent containing the tracking information, including the DCID, O-DCID, and the IP 5-tuple. Upon receiving a message, the API updates or creates a corresponding entry in the shared tracking table with the DCID as the primary key. The algorithm is summarized in Algorithm~\ref{algo:client_side_api}.

\begin{algorithm}[!ht]
\caption{Client-Facing API}
\begin{algorithmic}[1]
\State \textbf{Input:} dcid, o-dcid, ip\_tuple (from Client Agent)
\State \textbf{Shared:} Tracking table (dictionary)
\State Update tracking table with dcid, o-dcid, ip\_tuple
\end{algorithmic}
\label{algo:client_side_api}
\end{algorithm}


\paragraph{Middlebox-Facing API (Reactive Mode)}
The Middlebox-Facing API handles requests from middleboxes running in reactive mode. For each incoming query containing a DCID, it retrieves the corresponding tracking information from the shared table and responds with the associated O-DCID. If no entry exists, it returns \textit{None}. This operation is performed continuously in a dedicated thread, as illustrated in Algorithm~\ref{algo:middlebox_api}.

\begin{algorithm}[!ht]
\caption{Middlebox-Facing API (Reactive Mode)}
\begin{algorithmic}[1]
\State \textbf{Shared:} Tracking table (dictionary)
\State Initialize socket
\While{True}
\State Receive message from middlebox
\State Extract dcid, ip\_tuple
\State Retrieve o-dcid from tracking table
\State Package reply with tracking information
\State Send reply to middlebox
\EndWhile
\end{algorithmic}
\label{algo:middlebox_api}
\end{algorithm}

\paragraph{Middlebox-Facing API (Proactive Mode)}
In proactive mode, the Tracking Agent maintains a list of registered middleboxes for each QUIC flow. Whenever new tracking information becomes available, it is asynchronously pushed to all registered middleboxes. Algorithm~\ref{algo:middlebox_api_offline} formalizes this behavior.

\begin{algorithm}[!ht]
\caption{Middlebox-Facing API (Proactive Mode)}
\begin{algorithmic}[1]
\State \textbf{Shared:} Tracking table (dictionary)
\State \textbf{Shared:} Registered middleboxes list
\State Initialize socket
\Function{register}{address, quic\_ip\_tuple}
\State Add address to \texttt{registered\_middleboxes[quic\_ip\_tuple]}
\EndFunction
\Function{async\_push}{}
\While{True}
\If{New tracking information available}
\State Package tracking information
\State Retrieve registered middleboxes for flow
\For{each address in registered middleboxes}
\State Send update to address
\EndFor
\EndIf
\EndWhile
\EndFunction
\end{algorithmic}
\label{algo:middlebox_api_offline}
\end{algorithm}

\subsection{Middlebox}

We implemented the Default and proposed QUIC-aware Stateful Middlebox for RL and NAT using scapy~\cite{scapy_introduction}. Using this implementation, we inspected the service disruption caused by connection migration and evaluated our proposed framework. The implementation of the two QUIC-aware Stateful Middleboxes and their default counterparts is detailed in this section.

\subsubsection{Default NAT}
The default NAT configuration employs two network interfaces, designated as private and public, and supports a single public IP address. A Scapy-based packet sniffer operates on both interfaces to capture and process traffic from the private and public networks. Packets captured on the private interface are processed according to Algorithm \ref{algo:default_nat_private}, where the source IP and port are translated to the corresponding public IP and port before forwarding through the public interface. Conversely, packets captured on the public interface are processed as described in Algorithm \ref{algo:default_nat_public}; the private IP and port are retrieved from the connection tracking table, and the packet’s destination IP and port are updated accordingly before transmission through the private interface.

\begin{algorithm}[!ht]
    \caption{Processing Packets from the Private Interface by Default NAT}
    \begin{algorithmic}[1]
        \State \textbf{Input:} Packet arriving at private interface
        \State \textbf{Shared:} Connection tracking table
        \If{Packet protocol is not supported}
        \State Drop the packet
        \Else
        \State Extract 5-tuple (src IP, src port, dest IP, dest port, protocol)
        \State Lookup or create corresponding public address in tracking table
        \If{No public address assigned}
        \State Drop the packet
        \Else
        \State Update packet source IP and port to public address
        \State Forward packet through public interface
        \EndIf
        \EndIf
    \end{algorithmic}
    \label{algo:default_nat_private}
\end{algorithm}

\begin{algorithm}[!ht]
    \caption{Processing Packets from the Public Interface by Defauly NAT}
    \begin{algorithmic}[1]
        \State \textbf{Input:} Packet arriving at public interface
        \State \textbf{Shared:} Connection tracking table
        \If{Packet protocol is not supported}
        \State Drop the packet
        \Else
        \State Extract 5-tuple (src IP, src port, dest IP, dest port, protocol)
        \State Retrieve corresponding private address from tracking table
        \If{No private address found}
        \State Drop the packet
        \Else
        \State Update packet destination IP and port to private address
        \State Forward packet through private interface
        \EndIf
        \EndIf
    \end{algorithmic}
    \label{algo:default_nat_public}
\end{algorithm}

\subsubsection{QUIC-aware Stateful NAT}
The default NAT is extended to implement a QUIC-aware NAT that specifically handles QUIC traffic while processing other packets according to the default NAT configuration. In this NAT, QUIC packets are processed based on a 6-tuple consisting of the destination connection ID (DCID) and the standard 5-tuple (source IP, source port, destination IP, destination port, protocol).

Algorithm~\ref{algo:quic_nat_private} describes the procedure for processing QUIC packets arriving at the private interface. For each packet, the DCID is extracted and used to retrieve the corresponding tracking information from either the cached DCID table or the Tracking Agent. The public IP and port are obtained from the connection tracking table. The tracking table is updated with the latest private IP and port for the QUIC connection, replacing any previous entry. The packet is then updated and forwarded through the public interface. Incoming packets from the public interface are processed as in the default NAT since the connection tracking table already contains the updated private IP and port for the connection.

\begin{algorithm}[h!]
\caption{Processing Packets from the Private Interface by QUIC-aware NAT}
\begin{algorithmic}[1]
\State \textbf{Shared:} Connection tracking table, QUIC DCID tracking table
\If{Packet protocol is not QUIC}
\State Process the packet according to default NAT configuration
\Else
\State Extract ip\_6\_tuple (DCID + 5-tuple) from QUIC packet
\If{DCID exists in QUIC DCID table}
\State Retrieve tracking information from DCID table
\Else
\State Query Tracking Agent for tracking information
\If{No existing DCID found}
\State Initialize tracking information for new flow
\EndIf
\EndIf
\State Retrieve or create public address in connection tracking table
\State Update tracking table with current private IP and port
\If{Public address is None}
\State Drop the packet
\Else
\State Update packet source IP and port to public address
\State Forward packet through public interface
\EndIf
\EndIf
\end{algorithmic}
\label{algo:quic_nat_private}
\end{algorithm}

\subsubsection{Default Rate Limiter (RL)}
The default rate limiter implements flow-based token bucket rate limiting on packets traversing the network interfaces. Each flow is identified by its 5-tuple consisting of source IP, source port, destination IP, destination port, and protocol. A token bucket is maintained for every flow in the tracking table, with the capacity of each bucket increasing at a fixed rate up to a predefined maximum. When a packet is captured, the corresponding token bucket is decremented. If sufficient tokens are available, the packet is forwarded; otherwise, it is dropped. The detailed procedure is presented in Algorithm~\ref{algo:default_rl}.

\begin{algorithm}[!ht]
\caption{Flow-level Token Bucket Default RL}
\begin{algorithmic}[1]
\State \textbf{Shared:} Tracking table mapping flows to token buckets
\State Extract 5-tuple (protocol, src IP, dst IP, src port, dst port) from packet
\State Retrieve or create token bucket corresponding to the flow
\If{Token bucket has available tokens}
\State Forward the packet
\Else
\State Drop the packet
\EndIf
\end{algorithmic}
\label{algo:default_rl}
\end{algorithm}

\subsubsection{QUIC-aware Rate Limiter (RL)}
The default rate limiter is extended to support QUIC connection tracking, resulting in a QUIC-aware RL. For each QUIC packet, the destination connection ID (DCID) is extracted. Using the DCID, the corresponding original connection ID (O-DCID) is retrieved from either the local memory or the Tracking Agent. The RL maps the flow to a token bucket using the O-DCID and applies rate limiting as described in Algorithm~\ref{algo:quic_rl}.

For reverse flows, the DCID differs from that used by the client, so the flow is identified based on the packet’s destination IP and port. These fields are expected to exist in the tracking table, allowing the RL to correctly associate the packet with its flow. This ensures that QUIC connections are accurately rate-limited while preserving the semantics of the token bucket mechanism.

\begin{algorithm}[h!]
\caption{Flow-level Token Bucket QUIC-aware RL}
\begin{algorithmic}[1]
\State \textbf{Shared:} Connection tracking table mapping flows to token buckets, QUIC DCID table
\If{Packet protocol is not QUIC}
\State Process according to default RL configuration
\Else
\State Extract ip\_6\_tuple (DCID + 5-tuple) from QUIC packet
\If{DCID exists in QUIC DCID table}
\State Retrieve O-DCID from DCID table
\Else
\State Query Tracking Agent for O-DCID
\If{No O-DCID found}
\State Initialize O-DCID as DCID for new flow
\EndIf
\EndIf
\State Retrieve or create token bucket corresponding to the flow
\If{Token bucket has available tokens}
\State Forward the packet
\Else
\State Drop the packet
\EndIf
\EndIf
\end{algorithmic}
\label{algo:quic_rl}
\end{algorithm}

\end{document}